\documentstyle[pra,aps,twocolumn,graphics,epsfig]{revtex}
\begin{document}
\title{Classical and quantum signatures of competing $\chi^{(2)}$
nonlinearities}
\author{A.G. White, P. K. Lam, M. S. Taubman, M. A. M. Marte$^{*}$,
S. Schiller$^{\dagger}$, D.  E.  McClelland, \& H.-A.  Bachor}
\address{{Physics Department, Australian National University, Canberra 
ACT 0200, Australia}\\
Phone: +61(6) 2493523, Fax: +61(6) 2492747, 
e-mail: andrew.white@anu.edu.au \\
$^{*}$ Institut f\"ur Physik, Universit\"at Innsbruck, A-6020,
Innsbruck, Austria. \\
$^{\dagger}$ Fakult\"at f\"ur Physik, Universit\"at Konstanz, D78434,
Konstanz, Germany.}
\date{\small received 26 August 1996, accepted February 28 1997.}
\maketitle

\begin{abstract}
We report the first observation of the quantum effects of competing 
$\chi^{(2)}$ nonlinearities.  We also report new classical signatures 
of competition, namely clamping of the second harmonic power and 
production of nondegenerate frequencies in the visible.  Theory 
is presented that describes the observations as resulting from 
competition between various $\chi^{(2)}$ upconversion and 
downconversion processes.  We show that competition imposes hitherto 
unsuspected limits to both power generation and squeezing.  The 
observed signatures are expected to be significant effects in 
practical systems.\\
\\
PACS number(s): 42.50.-p, 42.65-k, 42.79.Nv, 03.65.Sq, 06.30.Ft \\
\end{abstract}

Second order, or $\chi^{(2)}$, nonlinear optical systems are employed 
successfully in applications ranging from frequency conversion to 
quantum optics.  The four basic $\chi^{(2)}$ processes are second 
harmonic and sum frequency generation (SHG and SFG, upconversion); and 
difference frequency generation and (non)degenerate optical parametric 
oscillation (DFG or (N)DOPO, downconversion).  In recent years there 
has been increasing interest in the behaviour of {\it interacting} 
$\chi^{(2)}$ nonlinearities.

Interacting nonlinearities can be categorised as {\it cooperating} and 
{\it competing}.  Cooperating nonlinearities are those where all the 
downconversion and upconversion processes share the same modes, e.g.  
$\nu \rightleftharpoons 2 \nu$ or $\nu \pm \Delta_{1} 
\rightleftharpoons 2 \nu$.  Competing nonlinearities are those where 
all the downconversion and upconversion processes do not share the 
same modes, e.g.  $\nu \rightleftharpoons 2 \nu \rightleftharpoons \nu 
\pm \Delta_{2}$, or, $\nu \pm \Delta_{1} \rightleftharpoons 2 \nu
\rightleftharpoons \nu \pm \Delta_{2}$).  Both forms of interaction
are often referred to as {\it cascaded} nonlinearities.

An early study of cooperating $\chi^{(2)}$ nonlinearities predicted 
power limiting of the pump in an optical parametric oscillator
\cite{Siegman}.  More recently the large third order effects possible 
via cooperating $\chi^{(2)}$ nonlinearities has been the subject of 
extensive research \cite{Assanto,Stegeman}, including CW studies using 
cavities \cite{KerrCW,Ou}.  Systems of cooperating nonlinearities hold 
promise for applications including optical switching, nonlinear 
optical amplification \cite{PatChu}, squeezing, and QND measurements 
\cite{KerrCW}.

In contrast, systems of competing nonlinearities have been mainly investigated 
for their potential as frequency tunable sources of light.  Systems 
considered include: intracavity SFG and NDOPO \cite{Moore93,Cheung94}; 
intracavity DFG and NDOPO \cite{Koch95}; and intracavity SHG and NDOPO
\cite{Byer,Schiller93,Marte94,Moore95,SchillerSPIE,SchillerAPL}.
The quantum properties of the latter system have been modelled for the 
quadruply resonant configuration \cite{MarteQuant} and several new 
nonclassical features are predicted.

In this paper we report the first experimental observation of the 
quantum effects of competing nonlinearities.  We also report two clear 
classical signatures of competition: power clamping of the second 
harmonic and production of nondegenerate optical frequencies in both 
the second harmonic and fundamental fields.

Figure 1 shows the conceptual layout.  A frequency doubler, resonant at and 
pumped by a frequency $\nu$, produces a nonresonant field of frequency $2 
\nu$ which is forced to make a double pass through the cavity.  The second 
harmonic can either downconvert back to the original mode, or act as the 
pump for the NDOPO.  For the latter to occur the signal \& idler modes 
($\nu_{s,i} = \nu \pm \Delta$) must be simultaneously resonant with the 
mode $\nu$.  With sufficient power in the $2 \nu$ field the NDOPO 
can be above threshold, otherwise the system is below threshold and acts as an 
amplifier of the vacuum modes.
\begin{figure}
\begin{center}
\epsfxsize=5cm
\epsfbox{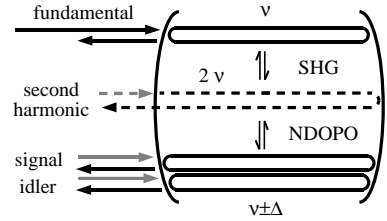}
\end{center}
\caption[Conceptual layout of TROPO.]{Conceptual layout of TROPO. Gray
lines represent vacuum inputs, i.e.  zero average power.}
 \label{fig:TROPO}
\end{figure}

The three equations of motion for this system are:
\begin{eqnarray}
\dot{\alpha}_{1} &=& -(\gamma_{1}+ i \Delta_{1}) \alpha_{1}
     - 2 \sqrt{ \mu_{1} \mu_{2} } \alpha_{1}^{*} \alpha_{s} \alpha_{i} 
     - \mu_{1} |\alpha_{1}|^{2} \alpha_{1} \nonumber \\
     & \phantom{=} &
     + \sqrt{ 2 \gamma_{1}^{\rm{c}} } A_{1} \nonumber \\
\dot{\alpha}_{s,i} &=& -(\gamma_{s,i} + i \Delta_{s,i}) \alpha_{s,i}
               - \sqrt{ \mu_{1} \mu_{2} } \alpha_{1}^{2} \alpha^{*}_{i,s}
               - 2 \mu_{2} \alpha_{s} \alpha_{i} \alpha^{*}_{i,s}  \nonumber \\             
\label{motion}
\end{eqnarray}
where $\alpha_{1}$, $\alpha_{s}$ , $\alpha_{i}$ are the fundamental, 
signal, and idler field amplitudes, respectively; $\gamma_{x}$ and 
$\Delta_{x}$ are respectively the decay rate and detuning of mode $x$; 
$\gamma_{1}^{\rm{c}}$ is the decay rate of the fundamental coupling 
mirror; $\mu_{1}$ and $\mu_{2}$ are the respective nonlinear 
interaction rates for SHG and NDOPO; and $\rm{A}_{1} = \sqrt{
\rm{P}_{1}/ (h \nu) }$, where $\rm{P}_{1}$ is the pump power, $h$ is 
Planck's constant, and $\nu$ is the fundamental frequency.

We define the term $\gamma^{eff}_{x} = (\gamma_{x}+ i \Delta_{x})$ as 
the effective decay rate of mode $x$.  To see why, consider the case 
of a singly resonant frequency doubler ($\mu_{2} = 0$).  Without loss 
of generality, we can assume that the pump rate, $A_{1}$, is real.  If 
the detuning, $\Delta_{1}$, is zero, then the field value, 
$\alpha_{1}$, is real.  It is clear that the value of $\alpha_{1}$ is 
limited by the total decay rate $\gamma_{1}$: if $\gamma_{1}$ is large 
then the absolute value of $\alpha_{1}$ will be small.  Now consider 
non-zero detuning: the value of $\alpha_{1}$ becomes complex and is 
limited by both the decay rate, $\gamma_{1}$, and the detuning, 
$\Delta_{1}$.  If the detuning is very large, then even when the decay 
rate is very small the absolute value of $\alpha_{1}$ will be small.  
Thus the linear phase shift, $i \Delta_{1}$, introduced by detuning 
leads to a reduction of intensity, and can be said to effectively 
increase the decay rate of the cavity.

For zero detunings, the threshold power for competition is:
\begin{equation}
P^{\rm{thr}}_{1} = h 2 \nu 
\frac{ \bar{\gamma} }   { \gamma_{1}^{\rm{c}} } 
\frac{ \gamma_{1}^{2} } { \sqrt{\mu_{1} \mu_{2}} }
\frac{1}{4} \left( 1 +
r \frac{ \bar{\gamma} } {\gamma_{1} } \right)^{2}
\label{thresh}
\end{equation}
where $\bar{\gamma}$\,=\,$\sqrt{\gamma_{s} \gamma_{i}}$ and 
$\rm{r}$\,=\,$\sqrt{\mu_{1}/ \mu_{2}}$.  We introduce the scaled power 
$\rm{N}$\,=\,$\rm{P}_{1}/\rm{P}^{\rm{thr}}_{1}$.  For the likely 
experimental optimum, $\gamma_{s}$\,=\,$\gamma_{i}$\,=\,$\gamma_{1}$, 
$\mu_{1}$\,=\,$\mu_{2}$, we define a minimum threshold power, 
$\rm{P}^{\rm{min}}_{1}$\,=\,$h (2 \nu) \gamma_{1}^{2}/(\eta \mu_{1})$ 
where the cavity escape efficiency is $\eta$\,=\,$\gamma_{1}^{\rm{c}}/
\gamma_{1}$.

Obviously the threshold can be altered by changing the nonlinearities.  
Experimentally this is achieved via {\it phase matching}: i.e.  
altering the phase match in the nonlinear crystal by changing the 
orientation or temperature \cite{KerrCW}.  The threshold can also be 
altered via {\it dispersion matching}.  That is, altering the laser 
frequency or cavity length so that the signal and idler modes are 
unable to be resonant with the fundamental.  This corresponds to large 
signal and idler detunings but zero fundamental detuning.  The altered 
threshold is then described by substituting absolute values of the 
effective decay rates, $|\gamma^{eff}_{x}|$, for all the decay rates 
in equation \ref{thresh}.

A detailed description of the experimental setup is given in 
\cite{mclnr}.  In brief, the system is driven by a miniature diode pumped 
Nd:YAG ring laser (Lightwave 122) that produces a single mode of 
wavelength 1064 nm.  A mode cleaning cavity acts as a low-pass filter 
to remove excess quadrature noise (both amplitude and phase) from the 
laser beam.  The output of this drives the nonlinear cavity, which is 
a 12.5 mm long MgO:$\rm{LiNbO_3}$ monolithic crystal with dielectric 
mirror coatings on the curved end faces (R=14.24 mm).  The monolith is 
singly resonant at the fundamental; the second harmonic executes a 
double-pass through the crystal (residual second harmonic transmitted 
through the high reflector end is referred to as ``single-pass'').  
The laser is locked to the monolith, and the mode cleaner is locked to 
the laser, via separate Pound-Drever locking schemes.  The second 
harmonic is accessed via a low-loss dichroic, the reflected 
fundamental is accessed via the Faraday isolator - both beams are sent 
to either a balanced-homodyne pair and/or an optical spectrum 
analyser.

The obvious signature of competition in this system is production of 
nondegenerate frequency modes (when N$>$1).  When the monolith is 
repeatedly scanned through resonance, these modes cause distorted 
cavity lineshapes.  The frequency of the modes is measured by 
inspecting the infrared field reflected from the monolith with a 
spectrometer.  The signal \& idler pair are found to be up to 31 nm 
from degeneracy ($\nu_{s,i}$\,=\,1033 nm,1095 nm).  The nondegeneracy 
is limited by phase matching, dispersion, and mirror bandwidth ($\sim$ 
40 nm centred at 1064 nm).  Figure 2 shows, for scanned operation 
\cite{WhiteKN}, the observed threshold power (curve a) and the 
single-pass and double-pass second harmonic power (curves b
\& c) as a function of the crystal temperature.  Note that the 
threshold curve has two minima: roughly corresponding to maxima in the 
double pass and single pass power, respectively.  In the latter case, 
even though minimal second harmonic is produced, the intracavity 
second harmonic field is large enough to pump the NDOPO.
\begin{figure}
\begin{center}
\epsfxsize=7cm
\epsfbox{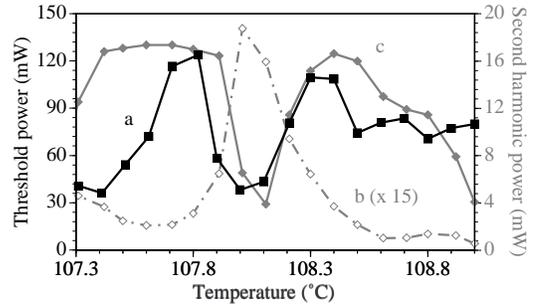}
\end{center}
\caption{(a) observed threshold power; (b) single-pass SH power
(i.e.  residual transmitted through high reflector) \& (c) double-pass 
SH power (as shown in Fig.  1); as a function of crystal temperature 
(i.e.  phase mismatch).}
 \label{fig:threshold vs temp}
\end{figure}

In locked operation the nondegenerate modes are observed via optical 
spectrum analysers.  Figure 3a is the output of the infrared optical 
spectrum analyser for the laser only.  Figure 3b is the output for the 
locked monolith just above threshold: note the strong conversion to 
signal and idler.  The signal and idler mode-hop irregularly, the 
system operating stably for up to ten minutes at a time.  Gross 
control is achieved by detuning the fundamental mode.  As it is 
detuned around resonance, the effective decay rate of the fundamental 
does not change greatly, but, due to dispersion mismatch, the 
effective decay rates of the signal and idler can become very large.  
This shifts the threshold power above the operating power and 
suppresses the NDOPO, c.f.  (\ref{thresh}).  Finer control has been 
achieved using a cavity with tunable dispersion, for example a 
semi-monolithic design where a translatable cavity mirror is external 
to the MgO:$\rm{LiNbO_3}$ crystal \cite{Schneider}.  Such improvements 
allow for stable operation with long intervals between mode hops.

Surprisingly, as the power is increased further two extra modes in the 
infrared, and four extra modes in the visible, are seen (Figures 
3c,d).  To the authors' knowledge this is the first observation of 
extra modes around the second harmonic, and it strongly supports the 
mechanism proposed in \cite{SchillerAPL} of cascaded second harmonic, 
sum and difference frequency generation between the signal, idler and 
pump fields.  The extra modes in the visible are likely generated by 
SFG ($\nu + \nu_{s,i} = 2 \nu \pm \Delta$) or SHG ($2 \nu_{s,i} = 2 
\nu \pm 2 \Delta$), whilst the extra pair in the infrared are from DFG 
with the visible modes (${ \nu +\nu_{s,i} } - \nu_{i,s} = 2 \nu_{s,i} 
- \nu = \nu \pm 2 \Delta$).  Further modes appear in the infrared 
field with increasing power: this system holds great promise both as a 
source of frequency tunable light and for frequency measurement (e.g.  
as a precise frequency chain).
\begin{figure}
\begin{center}
\epsfxsize=\columnwidth
\epsfbox{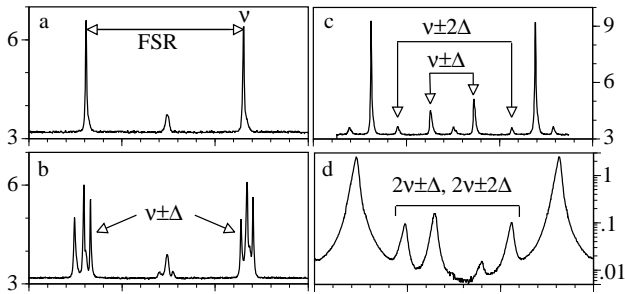}
\end{center}
\caption{Optical spectrum analyser outputs of the locked monolith. 
All traces are intensity versus frequency (arbitrary units).  The 
small peak in the middle of the infrared traces is due to imperfect 
alignment.  {\it infrared traces} (a) from laser for $\rm{P}_{1}=$32 
mW, FSR = free spectral range of the analyser; (b) from monolith for 
$\rm{P}_{1}=$14 mW, note signal and idler modes; (c) from monolith for 
$\rm{P}_{1}=$49 mW, note extra pair of modes; {\it visible trace} (d) 
from monolith for $\rm{P}_{1}=$155 mW , the ordinate is logarithmic to 
highlight the four extra frequencies.}
 \label{fig:OSA outputs}
\end{figure}

Another surprising signature of competition is clamping of the second 
harmonic power.  From (\ref{motion}) we find that for 
$P_{1}>P_{1}^{\rm{thr}}$, the second harmonic power is:
\begin{equation}
P_{2} = h 2 \nu \frac{ \bar{\gamma}^{2} } {\mu_{2}} 
\label{clamp}
\end{equation}
i.e the power is clamped to its threshold value.  Above threshold, 
``excess'' pump power is reflected or converted to signal and idler.  
Similar behaviour has been predicted for the {\it optical limiter} 
\cite{Siegman}: a standing wave DOPO resonant at $\nu$, which is 
single-pass pumped at $2 \nu$.  The $2 \nu$ field in both cases sees 
three input/output ports, however the clamping is due to different 
mechanisms: competing $\chi^{(2)}$ nonlinearities in our system; 
cooperating $\chi^{(2)}$ nonlinearities in the limiter.

The conversion efficiency at threshold is given by 
$\epsilon$\,=\,$P_{2}/P^{\rm{thr}}_{1}$.  The minimum threshold, 
$\rm{P}^{\rm{min}}_{1}$, is the point of maximum conversion 
efficiency, with a value equal to the cavity escape efficiency, 
$\epsilon$\,=\,$\gamma_{1}^{c}/\gamma_{1}$\,=\,$\eta$.  For unity 
cavity escape efficiency, $\eta=1$, $\rm{P}^{\rm{min}}_{1}$ is 
also the impedance matching point of the cavity.

Figure 4 shows experimental curves of second harmonic versus 
fundamental power for two different detunings.  In curve (a) the 
second harmonic power is clamped at 23 mW at a threshold power of 41 mW.  
This threshold is much higher than the observed minimum 
threshold, $\rm{P}^{\rm{min}}_{1}=14.3$ mW, as the signal 
and idler modes see high cavity losses due to dispersive mismatch.  In 
curve (b) the monolith is tuned towards resonance so that the 
effective fundamental decay rate is lower than in curve (a), however 
the detuning increases the dispersive mismatch, and thus 
$\gamma_{s,i}$, suppressing the NDOPO and moving the threshold to 54 
mW.
\begin{figure}
\begin{center}
\epsfxsize=7cm
\epsfbox{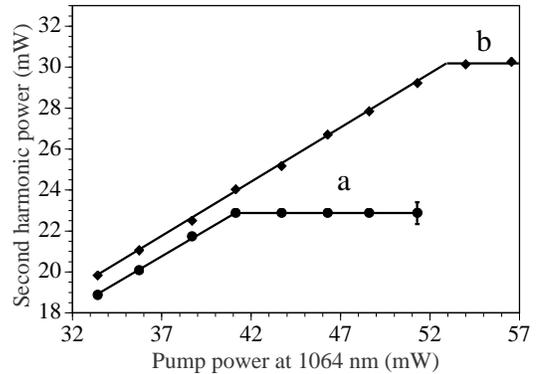}
\end{center}
\caption{Second harmonic power versus fundamental power curves for 
two different detunings, (a) \& (b). The systematic error bar is 
shown. All power measurements are NIST traceable with an absolute error 
of 7\%.}
 \label{fig:power clamp}
\end{figure}

This has important consequences when designing nonlinear optical 
systems.  Clamping is undesirable in many applications, such as 
frequency doubling to form a high power light source.  With the 
development of low dispersion, efficient nonlinear cavities, clamping 
is expected to become a widely observed phenomenon.  In the past year 
alone it has been observed in systems with competing SHG and NDOPO
\cite{WhiteLH,Schneider,Polzik} and in an optical limiter formed by an 
OPO intracavity with a laser \cite{Dunn}.  It can be suppressed via 
tunable dispersion, or avoided entirely by designing the system so 
that the minimum threshold point occurs at a power higher than maximum 
pump power.  Ideally clamping shouldn't occur in many frequency 
doublers as they are optimised for maximum conversion efficiency, i.e.  
pumped at $\rm{P}^{\rm{min}}_{1}$.  However in practice, many doublers 
are optimised by pumping them at powers above $\rm{P}^{\rm{min}}_{1}$.  
This is done because for powers less than $\rm{P}^{\rm{min}}_{1}$ the 
conversion efficiency falls off very steeply: small variations in 
fundamental power lead to large variations in harmonic power.  However 
above $\rm{P}^{\rm{min}}_{1}$ the conversion efficiency falls off very 
slowly: the harmonic power is much more robust to small variations in 
the fundamental power.  It is exactly this regime which is prone to 
competition.

Naturally, competition also has quantum signatures.  It has been 
suggested that, as the vacuum modes at $\nu_{s,i}$ are correlated by 
the NDOPO for N$<$1, then competition could be observed as squeezing 
of the reflected pump field at detection frequencies around the 
difference frequency of the signal and idler, $\Delta$ \cite{Polzik}.  
In our experiment the free spectral range of the monolith, (which sets 
the minimum value of $\Delta$), is much larger than the maximum 
bandwidth of the detectors (5.4 GHz and 100 MHz, respectively), ruling 
out any observation of this signature.

For the case where the second harmonic is resonant, the predicted 
quantum signature of competition is near perfect squeezing on either 
the fundamental or the second harmonic mode in power regimes that are 
inaccessible in the absence of competition \cite{MarteQuant}.  However 
in our system the second harmonic is not resonant, and the quantum 
signature of competition is very different: above threshold the 
squeezing degrades.  Without competition the second harmonic squeezing 
spectrum is given by
\cite{mclnr}:
\begin{eqnarray}
V_{2} &=& 1 - \frac{ 8 \gamma_{\rm{nl}}^{2} - 8\gamma_{\rm{nl}}
\gamma_{1}^{c} (V_{1}^{\rm{in}}-1) }
{ (3 \gamma_{\rm{nl}}^{2} + \gamma_{1})^{2} + \omega^{2} }
\label{normalsqz}
\end{eqnarray}
where the nonlinear loss rate, $\gamma_{nl}$\,=\,$\mu_{1} |\alpha_{1}|^2$; 
$\omega$\,=\,$2 \pi f$, where $f$ is the detection frequency; and 
$V_{1}^{\rm{in}}$ is the amplitude quadrature spectrum of the pump 
field.  For $\gamma_{\rm{nl}} >> \gamma_{1}$ and $V_{1}^{\rm{in}}$\,=\,$1$ 
the maximum squeezing of V=1/9 (-9.5 dB) is attained at zero frequency 
\cite{Paschotta}. With competing nonlinearities the spectrum becomes:
\begin{eqnarray}
V_{2} &=& 1 + 
\frac{2 (N -1) B(\omega) - 2 N A(\omega)}
{(N -1)^2 B(\omega) + \omega^2 (\frac{\gamma_{f}}{ 2 \bar{\gamma} })^2 +
  \frac{C(N) N A(\omega)} {r} + (\frac{\omega^2} {2 \bar{\gamma} } )^2 }
  \nonumber \\
\label{badsqz}
\end{eqnarray}
where N$>$1; $\gamma_{f}$\,=\,$\gamma_{1} + \rm{r} \bar{\gamma}$; 
$\rm{A}(\omega)$\,=\,$\rm{r}^{2} \omega^{2}$; 
$\rm{B}(\omega)$\,=\,$\gamma_{f}^{2} +\omega^{2}$; 
$\rm{C}(\rm{N})$\,=\,$\gamma_{1}/\bar{\gamma} + \rm{r} (\rm{N} +1) + 2 
(\rm{N} -1)$; and $V_{1}^{\rm{in}}$\,=\,$1$, $\gamma_{1}^{c} = 
\gamma_{1}$, for clarity.  If we assume the minimum threshold for 
competition, $\rm{P}^{\rm{min}}_{1}$, then 
$\gamma_{s}=\gamma_{i}=\gamma_{1}$ and $\mu_{1}=\mu_{2}$ and equation
\ref{badsqz} simplifies to:
\begin{equation}
V_{2} = 1 + \frac{2(N-1-\hat{\omega^{2}})}{4 N^{2} 
\hat{\omega^{2}} + (N-1-\hat{\omega^{2}})^{2}}
 \label{simplebadsqz}
\end{equation}
where $\hat{\omega}=\omega/(2 \gamma_{1})$.  A detailed theoretical 
discussion of the squeezing behaviour under these simplified 
conditions is given in \cite{Schiller97}.  Maximum squeezing occurs at 
the point where competition begins.  For the minimum threshold, 
$\rm{P}^{\rm{min}}_{1}$, the maximum squeezing is at zero frequency 
with a value V=1/2 (-3 dB).  For higher thresholds, 
$\rm{P}^{\rm{thr}}_{1} > \rm{P}^{\rm{min}}_{1}$, the maximum squeezing 
is still at zero frequency , but with larger values.  In all cases eqn
\ref{simplebadsqz} connects to eqn \ref{normalsqz} without 
discontinuity.

As Fig.  5 shows, for N$>$1 two effects come into in play, both of 
which degrade the squeezing.  Increasing N pulls the second harmonic 
noise, at all frequencies, towards the noise of the second harmonic 
input field.  As this is a vacuum field, the noise is pulled towards 
shot noise, regardless of whether it was originally above 
(super-Poissonian) or below (sub-Poissonian) shot noise.  Thus 
increasing N causes broadband degradation of the squeezing.  This 
behaviour is exactly analogous to that of an electro-optic noise 
eater, where increasing the beamsplitter reflectivity pulls the noise 
towards the limit set by the vacuum entering the empty beamsplitter 
port \cite{Taubman}.  This noise eating behaviour is expected to occur 
in other nonlinear optical systems: the optical limiter
\cite{Collett}; and the saturated laser amplifier \cite{Gray}.

The additional squeezing degradation evident at low frequencies is due 
to a more subtle effect.  In a conventional OPO, the signal \& idler 
amplitude quadratures are very noisy above threshold (for a DOPO the 
amplitude is shot noise limited at $\rm{P}=4 \rm{P}_{\rm{thr}}$ and 
50\% squeezed only for $\rm{P}>25 \rm{P}_{\rm{thr}}$ \cite{Slosser}).  
This noise is transmitted to the amplitude of the second harmonic, 
degrading the squeezing.  This low frequency degradation decreases 
with increasing N (compare curves b \& c in Fig. 5).
\begin{figure}
\begin{center}
\epsfxsize=7cm
\epsfbox{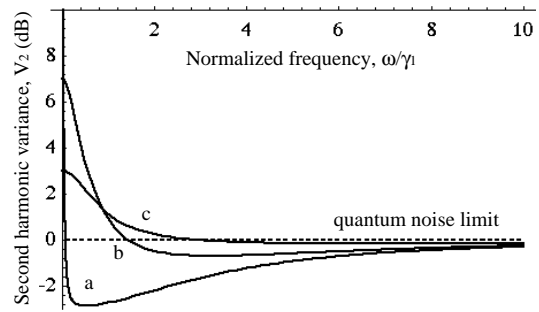}
\end{center}
\caption{Theoretical squeezing spectra for the case
$P^{\rm{thr}}_{1}$\,=\,$P^{\rm{min}}_{1}$.  (a) N$\,=\,$1.001 (b) 
N$\,=\,$1.25 and (c) N$\,=\,$3.}
 \label{fig:theoretical TROPO spectra}
\end{figure}
\begin{figure}
\begin{center}
\epsfxsize=7cm
\epsfbox{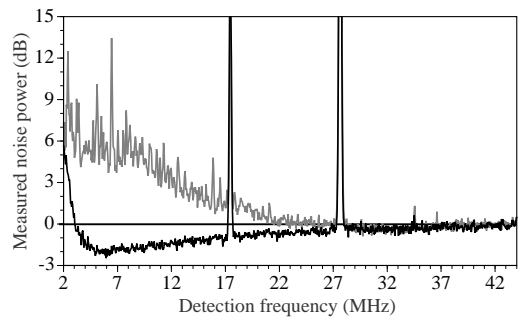}
\end{center}
\caption{(a) Squeezing spectra. (a) Without competition, $\rm{P}_{1}=74$ 
mW (b) With competition, $\rm{P}_{1}=60$ mW.}
 \label{fig:experimental TROPO spectra}
\end{figure}

Figures 6a shows the experimentally observed squeezing in the absence 
of competition, which is suppressed via detuning as discussed earlier.  
Below 6 MHz the squeezing degrades due to laser pump noise 
\cite{mclnr}, above 6 MHz the squeezing is as expected from theory 
with $V_{1}^{\rm{in}}=1$.  The spikes at 17 and 27 MHz are from the 
locking signals.  With competition, and at a lower pump power, the 
spectrum changes to that shown in Figure 6b.  As predicted, there is 
considerable excess noise at low frequencies, whilst the degradation 
at higher frequencies is more gradual.  The excess noise at low 
frequencies is greater than that shown in Fig.  5 due to presence of 
numerous, overlapping, noise spikes.  The spikes are due to a locking 
instability in the modecleaner which is driven by competing locking 
signals.  It is clear that even a small amount of $\chi^{2}$ 
competition leads to a marked degradation in the squeezing.  This 
previously unexpected limit to squeezing can only be avoided by 
designing the system so that competition is suppressed when the pump 
power is greater than the maximum conversion efficiency power.  One 
solution is a cavity with such high dispersion that the signal and 
idler modes are unable to become simultaneously resonant with the 
fundamental: high second harmonic squeezing has been seen in such a 
system \cite{Tsuchida}.

In conclusion, competition between SHG and NDOPO in a monolithic 
cavity has been observed to cause generation of new frequencies in 
both the visible and infrared fields, clamping of the second harmonic 
power, and degradation of the second harmonic squeezing.  Competition 
imposes a previously unsuspected limit to squeezing and power 
generation.  The reported signatures are expected to be commonly 
observed in efficient, low dispersion systems, unless explicit steps 
are taken to avoid competition.

We wish to acknowledge fruitful discussions with M. Collett, C. 
Savage, E. Polzik, J. Hall and B. Byer.  AGW wishes to thank J. Mlynek 
for the use of crystal \#19 for the unlocked measurements.  The ANU 
crystal was cut and polished by CSIRO, Sydney, Australia.  The 
coatings were provided by LZH, Hannover, Germany.  This work was 
supported by the Australian Research Council.  MM was supported by 
the APART programme of the Austrian Academy of Sciences.


\begin{thebibliography}{123}
\bibitem{WhiteKN} The scanned measurements were performed by AGW at Konstanz 
Universit\"at, Germany using crystal \#19.
\end{thebibliography}
\end{document}